\documentclass[twocolumn]{revtex4-1}
\usepackage{xcolor}
\usepackage{graphicx}
\usepackage{braket}
\usepackage{amsmath}
\usepackage{amssymb}

\newcommand{\wt}[1]{\widetilde{#1}}

\begin{document}

\title{Ensemble variational Monte Carlo for optimization of correlated excited state wave functions} 
\author{William A. Wheeler}
\affiliation{Department of Materials Science and Engineering; University of Illinois at Urbana-Champaign} 
\author{Kevin G. Kleiner}
\affiliation{The Anthony J Legget Institute for Condensed Matter Theory; Department of Physics; University of Illinois at Urbana-Champaign} 
\author{Lucas K. Wagner}
\affiliation{The Anthony J Legget Institute for Condensed Matter Theory; Department of Physics; University of Illinois at Urbana-Champaign} 
\begin{abstract}

Variational Monte Carlo methods have recently been applied to the calculation of excited states; however,
it is still an open question what objective function is most effective. 
A promising approach is to optimize excited states using a penalty to minimize overlap with lower eigenstates,
which has the drawback that states must be computed one at a time. 
We derive a general framework for constructing objective functions with minima at the the lowest $N$ eigenstates of a many-body Hamiltonian.
The objective function uses a weighted average of the energies and an overlap penalty, which must satisfy several conditions. 
We show this objective function has a minimum at the exact eigenstates for a finite penalty, and provide a few strategies to minimize the objective function.
The method is demonstrated using \textit{ab initio} variational Monte Carlo to calculate the degenerate first excited state of a CO molecule. 
\end{abstract}
\maketitle

\section{Introduction}

Quantum Monte Carlo (QMC) has been established as a highly accurate method for computing ground-states of \textit{ab initio} systems.\cite{foulkes_quantum_2001,luchow_quantum_2011,shulenburger_quantum_2013,wagner_discovering_2016,williams_direct_2020}
Progress has also been made towards efficiently computing accurate excited states in QMC;\cite{ blunt_krylov_2015, blunt_density_2017, choo_symmetries_2018, benavides_excitations_2022, otis_promising_2023}
however, there are limitations to the methods that have been applied to \textit{ab initio} systems so far.
A state-averaging approach has been used to simultaneously optimize a set of ground and excited states, where a common set of orbitals and Jastrow are used to construct all the states.\cite{filippi_absorption_2009}
In many systems, optimal orbitals for describing an excited state are quite different from the optimal ground-state orbitals, so large determinant expansions are required to accurately represent these states using a shared set of orbitals.
An optimization approach using variance of the local energy has been used to target excited states without state averaging;\cite{zhao_efficient_2016, shea_size_2017, pineda_flores_excited_2019, otis_optimization_2023} however, relying on the variance leads to difficulty optimizing to the correct minimum in some situations.\cite{cuzzocrea_variational_2020}
An alternative method avoids state averaging and variance optimization by penalizing the wave function's  overlap with a set of known eigenstates.\cite{Higgott2019variationalquantum, pathak_excited_2021} 
This method only optimizes one excited state at a time, requiring several expensive calculations in sequence.

We note that there has been some work that optimizes all states simultaneously\cite{shepard_double_2022, entwistle_electronic_2023} using a modification of the method in Refs.~\cite{Higgott2019variationalquantum, pathak_excited_2021}.
Additionally, a recently proposed method simultaneously optimizes excited states by representing the ensemble as a determinant of many-body wave functions, avoiding the need to choose an overlap penalty.  \cite{pfau_natural_2023}
In the approaches of Refs.~\cite{shepard_double_2022, pfau_natural_2023}, there is a degenerate minimum in which all wave functions are linear combinations of the eigenstates.\cite{theophilou_energy_1979} 
While in principle one could rediagonalize the Hamiltonian in the subspace, this is inconvenient in variational Monte Carlo (VMC) since the wave function parameterizations  are often nonlinear and the off-diagonal matrix elements of the Hamiltonian typically have high variance.
Finally, Ref.~\cite{entwistle_electronic_2023} includes an overlap penalty between states $i$ and $j$ only if $i < j$, effectively implementing the sequential optimization of Refs.~\cite{Higgott2019variationalquantum, pathak_excited_2021} in a single calculation by using separate, dynamic cost functionals for each state.

In this paper, we define a single, penalty-based cost functional to optimize an ensemble of states simultaneously.
For a finite, given value of penalty (no free parameters), the functional has the eigenstates as its global minimum, with Hessian eigenvalues all greater than 0. 
Using a weighted average of energies, as proposed in Ref.~\cite{gross_rayleigh_1988}, this cost functional yields the low energy eigenstates, and the Hamiltonian does not need to be rediagonalized in the final low-energy subspace.
We derive conditions on the optimization parameters that ensure the ensemble optimizes to the lowest-energy eigenstates.
As a practical demonstration, we optimize this cost functional to find excited states of a CO molecule.
We believe that strategies based on this functional are likely to be efficient and practical.

\section{Penalty-based cost functional}

\textbf{Summary and statement } We aim to find the lowest $N$ eigenstates of a time-independent Hamiltonian whose eigenstates $\ket{\Phi_i}$ satisfy $H\ket{\Phi_i} = E_i \ket{\Phi_i}$.
The minimum of the $N$ state functional of  $\lbrace \Psi_i \rbrace$ 
\begin{equation}\label{eq:cost}
O[\lbrace \Psi_i\rbrace] = \sum_i w_i E[\Psi_i] + \lambda \sum_{i<j} |S_{ij}|^2,
\end{equation}
where $E[\Psi_i] = \braket{\Psi_i | \hat H| \Psi_i}$ is the energy expectation, $S_{ij} = \braket{\Psi_i | \Psi_j}$ is the overlap, $w_i > 0$ are weights, and $\lambda >0$ penalizes wave function overlap.
If $w_i > w_{j}$ for all $i, j$  where $E_i < E_{j}$, and
\begin{align}\label{eq:lambda_max}
	\lambda > \max_{i<j} \left[ (E_j - E_i) \frac{w_iw_j}{w_i - w_j} \right],
\end{align}
then $O$ is minimized when $\Psi_i = \Phi_i$ are equal to the lowest $N$ eigenstates of the Hamiltonian $H$.
We prove this condition in sections \ref{sec:preliminaries}-\ref{sec:critical_lambda}.

\subsection{Preliminaries}\label{sec:preliminaries}

If the states $\{\Psi_i\}$ are orthogonal, a few properties are known.
An orthogonal ensemble satisfies the variational upper bound property\cite{gross_rayleigh_1988}
\begin{equation}
\sum_i w_i E[\Psi_i] \ge \sum_i w_i E[\Phi_i].
\label{eq:upper_bound}
\end{equation}
If the weights are all equal, every set of orthogonal states in the low-energy space $\mathcal{LE}_N = {\rm span}[\Phi_0, \ldots, \Phi_{N-1}]$ has the same cost,\cite{theophilou_energy_1979} resulting in a continuum of critical points.
More generally, if the weights in any subset $\{w_j\}$ are equal, then any rotation of the corresponding states $\{\Phi_j\}$ leaves the cost functional unchanged.
We therefore require the weights to be strictly decreasing between nondegenerate states, $w_i > w_{i+1}$ if $E_i \ne E_{i+1}$, to ensure that an orthogonal ensemble optimizes to the target ensemble $\lbrace\Phi_i\rbrace$.\cite{gross_rayleigh_1988}
Of course, within a degenerate subspace, any rotation of states leaves the energy unchanged and is an acceptable solution.

Now we turn to the case of $\{\Psi_i\}$ that are allowed to be nonorthogonal, and the minimization of $O$.
${\Psi_i}$ are orthogonal in the limit $\lambda\rightarrow\infty$.
At the opposite extreme, if $\lambda=0$, $O$ is minimized when all states are the ground state.
We will show that there is a critical value $\lambda_c$, above which $O$ is minimized by the ensemble of eigenstates $\{\Phi_i\}$.

We consider an ensemble $\{\Psi_i\}$ within the low-energy space $\mathcal{LE}_N$.
For any choice of $w_i$ and $\lambda$, the ensemble that minimizes $O$ is guaranteed to be in $\mathcal{LE}_N$,\cite{gross_rayleigh_1988} since any components outside $\mathcal{LE}_N$ can be rotated into $\mathcal{LE}_N$ without increasing the overlap term.
We thus can parameterize the ensemble states as linear combinations of eigenstates in $\mathcal{LE}_N$,
\begin{equation}\label{eq:define_states}
\Psi_i = \sum_{j=0}^{N-1} c_{ij} \Phi_j,
\end{equation}
where $c_{ii} = \sqrt{1-\sum_j |c_{ij}|^2}$ is determined by the normalization.
Our target ensemble $\lbrace \Phi_i \rbrace$ is represented by the parameters $c_{ij} = \delta_{ij}$.
In general, the cost functional with this parameterization is represented as
\begin{align}\label{eq:cost_function}
O &= \sum_{ik} w_i |c_{ik}|^2 E_k + \lambda \sum_{i<j} \left|\sum_k c_{ik}c_{jk}\right|^2.
\end{align}
$O$ is fully parameterized by $c_{ij}$ with $i\ne j$, which are free to vary within $|c_{ij}| \le 1$.

\subsection{Determining critical $\lambda_c$}\label{sec:critical_lambda}

\subsubsection{$\lambda_c$ for two-state system}\label{sec:critical_lambda_2}

\begin{figure}
    \includegraphics{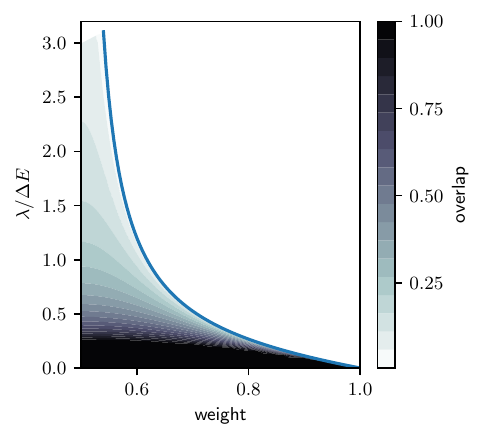}
    \caption{Overlap of the optimized ensembles in the two-state model system at different values of penalty $\lambda$ and weight $w_0$. The weights are normalized so that $w_0 + w_1 = 1$. The blue line is the calculated critical penalty $\lambda_c$, above which the overlap is strictly zero. 
}
\label{fig:lambda_vs_weight}
\end{figure}

We will determine the critical $\lambda_c$ such that the eigenstates $\lbrace \Phi_0, \Phi_1\rbrace$ minimize $O$ when $\lambda > \lambda_c$.
The cost functional for these two states is
\begin{align}
O &= \sum_{i=0}^{N-1} w_i E_i + w_0|c_{01}|^2 \Delta E - w_1 |c_{10}|^2 \Delta E 
\\&\qquad + \lambda |c_{00}c_{10} + c_{01}c_{11}|^2,
\end{align}
where $\Delta E = E_1 - E_0$.
The gradient and Hessian of $O$ give conditions for the critical penalty $\lambda_c$.
The gradient is
\begin{align}
\frac{\partial O}{\partial c_{01}}
&= 2 w_0 c_{01}\Delta E + \lambda 2 S_{01}\left(c_{11}-\frac{c_{01}c_{10}}{c_{00}}\right),
\end{align}
using the fact that $\frac{d c_{00}}{d c_{01}} = -c_{01}/c_{00}$.
In this form, it is clear that our target ensemble $c_{ij} = \delta_{ij}$ is indeed a critical point.
The Hessian matrix at the target ensemble $c_{ij}=\delta_{ij}$ is 
\begin{equation}
H_O =
\begin{pmatrix}
2 w_0\Delta E + 2\lambda & 2\lambda \\
2\lambda & -2 w_1\Delta E + 2\lambda
\end{pmatrix},
\end{equation}
where the rows and columns are the free parameters $ c_{01}$ and $c_{10}$.
It has eigenvalues
\begin{equation}
\alpha = 2\lambda + (w_0 - w_1)\Delta E \pm \sqrt{(w_0 + w_1)^2\Delta E^2 + 4\lambda^2}.
\end{equation}
This critical point is a minimum when the eigenvalues are both positive, which constrains $\lambda$ to
\begin{align}
\lambda &> \lambda_c = \Delta E \frac{w_0w_1}{w_0 - w_1}.
\label{eq:two_state_condition}
\end{align}

From the denominator, we see that if the weights are equal, $w_0 = w_1$, the critical $\lambda_c \rightarrow\infty$; in this case, no choice of $\lambda$ will minimize $O$ at the target ensemble.
In the other limit $w_1\rightarrow 0$, $\lambda_c\rightarrow 0$.
We expect the optimal choice of weights and $\lambda$ to be intermediate for efficient optimization.

We verify the condition of Eq.~\ref{eq:two_state_condition} by optimizing the cost functional in a two-state Hilbert space using different values of $\lambda$ and weights $(w_0, w_1)$.
The results are shown in Fig.~\ref{fig:lambda_vs_weight}.
The penalty $\lambda/\Delta E$ is plotted against the first weight $w_0$, and the weights are normalized, $w_0+w_1=1$.
The blue line shows the critical condition $\lambda_c / \Delta E = w_0w_1/(w_1 - w_0)$.
As expected, the overlap is zero when $\lambda > \lambda_c$, and is nonzero when $\lambda < \lambda_c$.
As $\lambda\rightarrow 0$, the overlap goes to one, as both states optimize to the ground state.

\subsubsection{$\lambda_c$ for $N$-state system}\label{sec:critical_lambda_N}

For an ensemble of $N$ states, the cost functional has the form given in Eq.~\eqref{eq:cost_function}.
We set $c_{ii}^2 = 1 - \sum_{j} |c_{ij}|^2$.
The gradients take the form
\begin{align} 
\frac{\partial O}{\partial c_{ij}} 
&= 2 w_i c_{ij} (E_{j} - E_i) + \sum_{k\ne i} 2\lambda S_{ik}\left(c_{kj} -\frac{c_{ij}c_{ki}}{c_{ii}}\right).
\end{align}
The derivative is zero when $\Psi_i = \Phi_i$, since $c_{ij}=\delta_{ij}$, and $S_{ij}=\delta_{ij}$.

We now check that the Hessian matrix evaluated at the target ensemble state has positive eigenvalues.
The Hessian matrix is made up of the second derivatives $\partial^2 O / \partial c_{ij} \partial c_{kl}$, where $i\ne j$ and $k\ne l$.
We start by identifying the nonzero off-diagonal terms, where $(i, j) \ne (k, l)$, 
\begin{align} 
H_O
&= 
2\lambda \sum_{m\ne i} \Bigg[ \frac{\partial S_{im}}{\partial c_{kl}} \left(c_{mj} -\frac{c_{ij}c_{mi}}{c_{ii}}\right)
\\&\qquad\qquad + 
S_{im}\frac{\partial}{\partial c_{kl}} \left(c_{mj} -\frac{c_{ij}c_{mi}}{c_{ii}}\right)\Bigg]
\\&= 2\lambda \sum_{m\ne i} \frac{\partial S_{im}}{\partial c_{kl}} \delta_{mj} 
\\&= 2\lambda \frac{\partial}{\partial c_{kl}} \sum_n c_{in} c_{jn} 
\\&= 2\lambda (\delta_{ik} c_{jl} + \delta_{jk} c_{il})
,
\end{align}
where we are at the target ensemble, $c_{ij} = \delta_{ij}$.
We see that the only nonzero off-diagonal term is $\partial^2 O / \partial c_{ij} \partial c_{ji}$, where $(k, l) = (j, i)$.
These off-diagonal terms are the result of two-state mixing, the situation described in the previous section.
Thus, the critical $\lambda_c$ satisfies the condition
\begin{align}\label{eq:lambda_c}
\lambda_c = \max_{i<j} \left[ (E_j - E_i) \frac{w_iw_j}{w_i - w_j} \right],
\end{align}
and when $\lambda > \lambda_c$, the target ensemble is a minimum of $O$.

\subsection{Choosing weights}
\subsubsection{Equal critical $\lambda$}\label{sec:choosing_weights}

In the sections above, we derived the critical value $\lambda_c$ that ensures the cost functional is minimized by eigenstates given a set of weights $\lbrace w_i\rbrace$.
As defined in Eq.~\ref{eq:lambda_c}, $\lambda_c$ is the maximum of the critical values for each pair of states.
One possible choice of weights is to fix all the pairwise critical values to be equal.
This strategy has one free parameter, the critical value itself.

Consider the ratios $\wt{w}_i = w_i / w_0$ and the energy differences $\wt{E}_i = E_i - E_0$.
The critical $\lambda_c^{i0}$ for states $i$ and $0$ are expressed as 
\begin{align}\label{eq:lambda_i0}
\frac{\lambda_c^{i0}}{w_0} = \wt{E}_i \frac{\wt{w}_i}{1 - \wt{w}_i}.
\end{align}
If we set all the $\lambda_c^{i0}$ equal to a chosen constant $\lambda_c$, the weight ratios are given by
\begin{align}\label{eq:w_tilde}
\wt{w}_i = \frac{1}{1 + \wt{E}_i w_0 / \lambda_c}.
\end{align}

With the weights determined this way, the critical $\lambda_c^{ij}$ for any pair $(i, j)$ also have the same value $\lambda_c$,
\begin{align}\label{eq:prove_lambda_same}
\frac{\lambda_c^{ij}}{w_0} 
&= (\wt{E}_i - \wt{E}_j) \frac{\wt{w}_i\wt{w}_j}{\wt{w}_j - \wt{w}_i} \\
&= (\wt{E}_i - \wt{E}_j) \frac{1}{1/\wt{w}_i - 1/\wt{w}_j} \\
&= (\wt{E}_i - \wt{E}_j) \frac{1}{1+\wt{E}_i w_0/\lambda_c - (1+\wt{E}_j w_0/\lambda_c)} \\
&= \frac{\lambda_c}{w_0}.
\end{align}

The value of $w_0$ is determined by normalizing the sum of the weights to one,
\begin{equation}
w_0 = \left(\sum_i \wt{w}_i \right)^{-1},
\end{equation}
and can be solved for given a desired $\lambda_c$ and an estimate of the energies $E_i$.
In practice, the energies $E_i$ must be estimated by an approximate method, since the optimized energies are not yet known.
The weights affect the efficiency of the optimization, but not the optimal solution, so exact energies are not necessary.
Note that it is not necessary for the weights to be normalized; however, only the ratio of the weights to $\lambda$ impacts the cost function, so we normalize the weights for interpretability.

\begin{figure}
\includegraphics{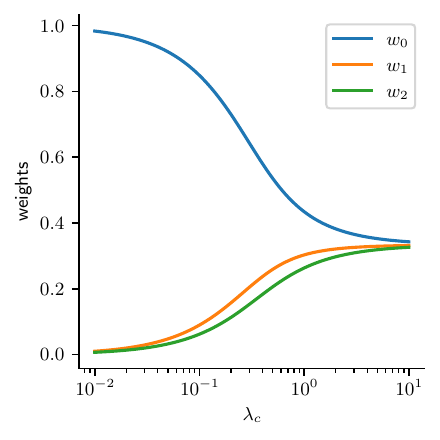}
\caption{Weights generated by the method in Sec.~\ref{sec:choosing_weights} at different values of $\lambda_c$. The energies in this example are $E = (0.0, 1.0, 1.5)$.}
\label{fig:weights_vs_lambdac}
\end{figure}

As an example, we show the weights generated by this method for a range of values of $\lambda_c$ in Fig.~\ref{fig:weights_vs_lambdac}.
At the extreme of small $\lambda_c$, $w_0$ is close to one, and all other weights are near zero.
This has the obvious problem that the excited states will not be well distinguished by the cost functional and will be dominated by stochastic noise from the ground state energy.
At the other extreme of large $\lambda_c$, the weights tend towards being equal.
As discussed above, equal weights cause the cost functional to have a degenerate minimum, where any rotation of the first $N$ eigenstates has the same energy as the eigenstates themselves. 
This effect is expected once noise in the overlap term (magnified by the large $\lambda$) dominates the differences in the weighted energies, even when the weights are not exactly equal.
In practice, we find that even deterministic optimization on simple models has difficulty reaching the minimum at these extremes of $\lambda_c$.

\subsubsection{Simultaneous `stack of states'}

Another strategy for optimization is to simultaneously optimize all states using the cost functions of Refs.~\cite{Higgott2019variationalquantum, pathak_excited_2021}, where each state $\Psi_j$ is orthogonalized only to lower states $\Psi_i$, $i < j$.
This strategy, used in Ref.~\cite{entwistle_electronic_2023}, is equivalent to the limit of a certain choice of weights in Eq.~\ref{eq:cost}, which we derive in this section.

Consider setting 
\begin{equation}
	\lambda_{ij} = \alpha_{ij} w_j,
\end{equation}
keeping in mind that the overlap sum in Eq.~\ref{eq:cost} is only over $i<j$, and where  $\alpha_{ij}$ is a weighting variable chosen so that  
\begin{equation}\alpha_{ij} > (E_j - E_i) /(1-w_j/w_i).\end{equation}
Then $\lambda_{ij} > \lambda_c$. 
The parameter derivative with respect to the parameter set $p_j$ of wave function $j$ is then 
\begin{align}
	\nabla_j O &= w_j \left( \nabla_j E[\Psi_j] +  \sum_{i<j} 2\alpha_{ij} S_{ij} \nabla_j S_{ij}   \right) \\ 
	&+ \sum_{k > j} 2 w_k \alpha_{jk} S_{jk} \nabla_j S_{jk}.
\end{align}

Now take the limit as $w_j / w_i \rightarrow 0$ while keeping the sum of weights equal to 1. 
This satisfies the ensemble criterion $w_i>w_j$ for $E_i < E_j$ and $w_i > 0$
and in this limit, $\alpha_{ij}$ only needs to be greater than $E_j - E_i$.
The gradients then become 
\begin{align}
	\nabla_j O &= w_j \left( \nabla_j E[\Psi_j] +  \sum_{i<j} 2\alpha_{ij} S_{ij}\nabla_j S_{ij}  \right),
\end{align}
which is equivalent to the algorithm recently proposed in Ref.~\cite{entwistle_electronic_2023}. 
In this version of the algorithm, the parameter updates can be made completely independent of the weights by scaling them as
\begin{equation}
	\Delta p_j = \frac{\tau}{w_j} F^{-1} \nabla_j O,
\end{equation}
where $F$ is a Fisher information matrix\cite{natural_gradient, sorella_green_1998}  and $\tau$ is the descent step length that can be optimized using correlated sampling\cite{pyqmc} or reduced on a schedule.

We have shown that a practical algorithm proposed in a heuristic manner actually rigorously converges to the eigenstates of the system. 
One should note that this choice of $\lambda_{ij}$ hides the upper bound property of $O$ (Eq.~\ref{eq:upper_bound}), making comparisons between different wave function ansatzes challenging. 
However, one can choose to evaluate any objective function on the optimized wave functions with finite weights and $\lambda_{ij}$ satisfying our $\lambda_c $ criterion to recover the upper bound property.

\section{Application to \textit{ab initio} system}

To demonstrate our penalty-based ensemble optimization on a simple, well-characterized system, we apply it to a CO molecule at the experimentally-determined equilibrium bond length of $r = 2.13$ Bohr \cite{tobiasPotentialEnergyCurves1960}.
We compute the ground state and degenerate first two excited states using VMC with different values of the overlap penalty $\lambda$, verifying the vertical excitation energy of absorption against experiment \cite{tobiasPotentialEnergyCurves1960}.

We represent our \textit{ab initio} wave functions using a multi-Slater-Jastrow ansatz,
\begin{equation}
\Psi(\textbf{R}) = e^{J(\textbf{R}; \alpha)}\sum_kc_kD_k(\textbf{R}; \beta),
\label{wf_ansatz}
\end{equation}
where $\textbf{R}$ represents all electron coordinates, $J$ is a two-body Jastrow factor with parameters $\alpha$, $D_k$ are Slater determinants of one-body orbitals parameterized by $\beta$, and $c_k$ are determinant expansion coefficients.
The determinants and expansion coefficients were generated from a complete active space configuration interaction (CASCI) calculation with six electrons and six orbitals, giving a basis of 400 determinants. 
The orbitals were computed using Hartree Fock (HF) with correlation-consistent effective core potentials and corresponding triple-$\zeta$ basis functions.\cite{bennet_new_2017}
The HF and CASCI calculations were carried out in PySCF, and the wave functions' Jastrow, determinant, and orbital parameters were optimized in PyQMC.

State 0 is initialized as the CASCI ground state multiplied by the Jastrow from a prior ground state optimization of Jastrow parameters in a multi-Slater-Jastrow wave function.
States 1 and 2 are initialized as equal-weight linear combinations of the CASCI ground state and target excited state multiplied by the same Jastrow as state 0.
We start states 1 and 2 at superpositions of CASCI eigenstates to avoid saddle points and ensure the method converges to the correct  excited states, even starting far from them.
The weights $\lbrace w_i\rbrace$ are set to $(0.41, 0.29, 0.29)$ using Eq.~\ref{eq:w_tilde} with $\lambda_c = 0.27$ Ha, the energy gap from the CASCI calculations.
All Jastrow parameters $\alpha$, orbital parameters $\beta$, and determinant coefficients $c_k$ are optimized with 500 walkers.
Vertical excitation energies are evaluated from the optimized wave functions using VMC with 10,000 walkers.

\begin{figure}
\includegraphics{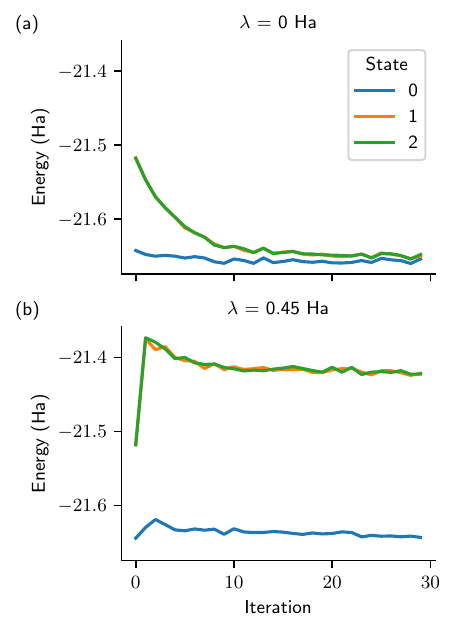}
\caption{Optimization of the objective functional Eq.~\ref{eq:cost_function} for the lowest three states of a CO molecule with (a) $\lambda = 0$ Ha or (b) $\lambda = 0.45$ Ha.} 
\label{fig:CO_lambda_comparison}
\end{figure}

Ensemble optimization using our cost functional generates the expected solutions in the limiting values of $\lambda$.
When $\lambda = 0$ Ha, all three states optimize to the ground state, shown in the left plot of Fig.~\ref{fig:CO_lambda_comparison}.
All three energies decrease over the optimization and approach the same value.
Our estimate for the critical penalty based on CASCI energies is $\lambda_c = 0.27$. 
When $\lambda = 0.45$ Ha, substantially larger than $\lambda_c$, state 0 optimizes to the ground state and states 1 and 2 optimize to the lowest two excited states, shown in in the right plot of Fig.~\ref{fig:CO_lambda_comparison}.
The energies of states 1 and 2, known to be degenerate, approach the same value of $0.220(5)$ Ha above the state 0 energy, even though they were initialized to a state much lower in energy.
The results confirm that the optimization works as expected on this system.
The VMC-calculated vertical excitation energy is within error bars of the experimentally determined $X^1\Sigma^+ \rightarrow a^3\Pi_r$ value of 0.22 Ha \cite{tobiasPotentialEnergyCurves1960}.

\begin{figure}
\includegraphics{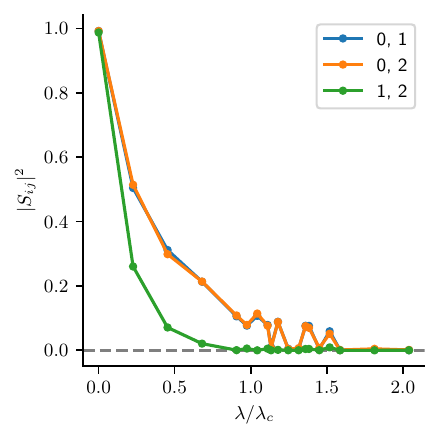}
\caption{Overlap between three wave functions of the CO molecule as a function of the overlap penalty $\lambda$. As the penalty approaches the critical value $\lambda_c$, the overlaps go to zero. An overlap larger than the critical value is necessary to stabilize the stochastic optimization. A lighted dashed line is shown at zero overlap for reference.}
\label{fig:CO_final_overlaps}
\end{figure}

We also confirm that our ensemble optimization yields the expected results for small and large values of $\lambda$.
As $\lambda$ increases from zero, the optimized states start out perfectly overlapping, and end up mutually orthogonal when $\lambda/\lambda_c$ is large, shown in Fig.~\ref{fig:CO_final_overlaps}
We find that orthogonality is only consistently achieved when $\lambda/\lambda_c>1.5$, although our derived condition only requires $\lambda/\lambda_c > 1$.
We attribute the difference to difficulty optimizing in the region $1 < \lambda/\lambda_c < 1.5$.
In this region, the overlaps are zero for some points and not others, a sign of inconsistent optimization.
When $\lambda/\lambda_c = 1$, we showed that the Hessian of the cost functional has a zero eigenvalue, meaning the cost functional is flat along at least one direction at the minimum.
For $\lambda/\lambda_c$ just above 1, this direction has very low curvature, making the minimum difficult to distinguish within stochastic noise.
For large enough $\lambda/\lambda_c$, the curvature of the cost functional along this overlap direction is higher, and the minimum is reliably found.

\section{Conclusion} 

We have presented an cost functional for simultaneously optimizing an ensemble of states to the low-energy eigenstates of a Hamiltonian in VMC.
Our approach is state-specific, allowing all parameters of each ensemble state to vary independently without any restrictions to enforce orthogonality.
The cost functional uses a weighted sum of energy expectations, distinguishing it from other recent approaches to ensemble optimization.
Whether the cost functional has the correct minimum (at the low energy eigenstates) depends on the choice of $\lambda$ and the weights.
In particular, we showed that there is a critical value of the penalty $\lambda_c$, which $\lambda$ must exceed; otherwise, the states that minimize the cost functional will not be orthogonal, and therefore not the correct eigenstates.
Finally, we demonstrated the proposed cost functional on an \textit{ab initio} VMC calculation of a carbon monoxide molecule.
We verified the expected behavior when the penalty $\lambda$ is below and above the estimated  $\lambda_c$, and that the method reaches the correct energy eigenstates even when initialized far from them.

We showed that there is a family of numerical methods which can satisfy the constraints noted in Eq.~\ref{eq:lambda_c} and Eq.~\ref{eq:cost_function}, at least one of which has been proposed in the literature without proof in the context of neural network wave functions.\cite{entwistle_electronic_2023} 
It is not clear which in this family are most optimal; an interesting future direction will be to determine the most convenient formulation of excited state optimization.
For all these techniques, the total computational cost of the excited state optimization is approximately $N$ times the cost of a single grounds state calculation, where $N$ is the number of excited states sought.
We anticipate the cost functional presented here to be useful in future excited state quantum Monte Carlo calculations with a variety of wave functions, including traditional as well as neural network and other machine-learning ansatzes.

\section{Acknowledgements}

This work was supported by U.S. National Science Foundation via Award No. 1931258 (WAW, LKW) and by the National Science Foundation Graduate Research Fellowship Program under Award No. DGE-1746047 (KGK).
\textit{Ab initio} calculations used computing resources from the Illinois Campus Cluster as well as the Flatiron Institute's Scientific Computing Core.
We would like to thank Kieron Burke for bringing ensemble DFT to our attention, Robert Webber and Michael Lindsey for early conversations, and Claudia Filippi for many comments and discussions.

\bibliography{ref.bib}

\begin{thebibliography}{28}%
\makeatletter
\providecommand \@ifxundefined [1]{%
 \@ifx{#1\undefined}
}%
\providecommand \@ifnum [1]{%
 \ifnum #1\expandafter \@firstoftwo
 \else \expandafter \@secondoftwo
 \fi
}%
\providecommand \@ifx [1]{%
 \ifx #1\expandafter \@firstoftwo
 \else \expandafter \@secondoftwo
 \fi
}%
\providecommand \natexlab [1]{#1}%
\providecommand \enquote  [1]{``#1''}%
\providecommand \bibnamefont  [1]{#1}%
\providecommand \bibfnamefont [1]{#1}%
\providecommand \citenamefont [1]{#1}%
\providecommand \href@noop [0]{\@secondoftwo}%
\providecommand \href [0]{\begingroup \@sanitize@url \@href}%
\providecommand \@href[1]{\@@startlink{#1}\@@href}%
\providecommand \@@href[1]{\endgroup#1\@@endlink}%
\providecommand \@sanitize@url [0]{\catcode `\\12\catcode `\$12\catcode
  `\&12\catcode `\#12\catcode `\^12\catcode `\_12\catcode `\%12\relax}%
\providecommand \@@startlink[1]{}%
\providecommand \@@endlink[0]{}%
\providecommand \url  [0]{\begingroup\@sanitize@url \@url }%
\providecommand \@url [1]{\endgroup\@href {#1}{\urlprefix }}%
\providecommand \urlprefix  [0]{URL }%
\providecommand \Eprint [0]{\href }%
\providecommand \doibase [0]{http://dx.doi.org/}%
\providecommand \selectlanguage [0]{\@gobble}%
\providecommand \bibinfo  [0]{\@secondoftwo}%
\providecommand \bibfield  [0]{\@secondoftwo}%
\providecommand \translation [1]{[#1]}%
\providecommand \BibitemOpen [0]{}%
\providecommand \bibitemStop [0]{}%
\providecommand \bibitemNoStop [0]{.\EOS\space}%
\providecommand \EOS [0]{\spacefactor3000\relax}%
\providecommand \BibitemShut  [1]{\csname bibitem#1\endcsname}%
\let\auto@bib@innerbib\@empty
\bibitem [{\citenamefont {Foulkes}\ \emph {et~al.}(2001)\citenamefont
  {Foulkes}, \citenamefont {Mitas}, \citenamefont {Needs},\ and\ \citenamefont
  {Rajagopal}}]{foulkes_quantum_2001}%
  \BibitemOpen
  \bibfield  {author} {\bibinfo {author} {\bibfnamefont {W.~M.~C.}\
  \bibnamefont {Foulkes}}, \bibinfo {author} {\bibfnamefont {L.}~\bibnamefont
  {Mitas}}, \bibinfo {author} {\bibfnamefont {R.~J.}\ \bibnamefont {Needs}}, \
  and\ \bibinfo {author} {\bibfnamefont {G.}~\bibnamefont {Rajagopal}},\ }\href
  {\doibase 10.1103/RevModPhys.73.33} {\bibfield  {journal} {\bibinfo
  {journal} {Rev. Mod. Phys.}\ }\textbf {\bibinfo {volume} {73}},\ \bibinfo
  {pages} {33} (\bibinfo {year} {2001})}\BibitemShut {NoStop}%
\bibitem [{\citenamefont {L\"uchow}(2011)}]{luchow_quantum_2011}%
  \BibitemOpen
  \bibfield  {author} {\bibinfo {author} {\bibfnamefont {A.}~\bibnamefont
  {L\"uchow}},\ }\href {\doibase 10.1002/wcms.40} {\bibfield  {journal}
  {\bibinfo  {journal} {WIREs Computational Molecular Science}\ }\textbf
  {\bibinfo {volume} {1}},\ \bibinfo {pages} {388} (\bibinfo {year}
  {2011})}\BibitemShut {NoStop}%
\bibitem [{\citenamefont {Shulenburger}(2013)}]{shulenburger_quantum_2013}%
  \BibitemOpen
  \bibfield  {author} {\bibinfo {author} {\bibfnamefont {L.}~\bibnamefont
  {Shulenburger}},\ }\href {\doibase 10.1103/PhysRevB.88.245117} {\bibfield
  {journal} {\bibinfo  {journal} {Physical Review B}\ }\textbf {\bibinfo
  {volume} {88}} (\bibinfo {year} {2013}),\
  10.1103/PhysRevB.88.245117}\BibitemShut {NoStop}%
\bibitem [{\citenamefont {Wagner}\ and\ \citenamefont
  {Ceperley}(2016)}]{wagner_discovering_2016}%
  \BibitemOpen
  \bibfield  {author} {\bibinfo {author} {\bibfnamefont {L.~K.}\ \bibnamefont
  {Wagner}}\ and\ \bibinfo {author} {\bibfnamefont {D.~M.}\ \bibnamefont
  {Ceperley}},\ }\href {\doibase 10.1088/0034-4885/79/9/094501} {\bibfield
  {journal} {\bibinfo  {journal} {Reports on Progress in Physics}\ }\textbf
  {\bibinfo {volume} {79}},\ \bibinfo {pages} {094501} (\bibinfo {year}
  {2016})}\BibitemShut {NoStop}%
\bibitem [{\citenamefont {Williams}\ \emph {et~al.}(2020)\citenamefont
  {Williams}, \citenamefont {Yao}, \citenamefont {Li}, \citenamefont {Chen},
  \citenamefont {Shi}, \citenamefont {Motta}, \citenamefont {Niu},
  \citenamefont {Ray}, \citenamefont {Guo}, \citenamefont {Anderson},
  \citenamefont {Li}, \citenamefont {Tran}, \citenamefont {Yeh}, \citenamefont
  {Mussard}, \citenamefont {Sharma}, \citenamefont {Bruneval}, \citenamefont
  {Van~Schilfgaarde}, \citenamefont {Booth}, \citenamefont {Chan},
  \citenamefont {Zhang}, \citenamefont {Gull}, \citenamefont {Zgid},
  \citenamefont {Millis}, \citenamefont {Umrigar}, \citenamefont {Wagner},\
  and\ \citenamefont {{Simons Collaboration on the Many-Electron
  Problem}}}]{williams_direct_2020}%
  \BibitemOpen
  \bibfield  {author} {\bibinfo {author} {\bibfnamefont {K.~T.}\ \bibnamefont
  {Williams}}, \bibinfo {author} {\bibfnamefont {Y.}~\bibnamefont {Yao}},
  \bibinfo {author} {\bibfnamefont {J.}~\bibnamefont {Li}}, \bibinfo {author}
  {\bibfnamefont {L.}~\bibnamefont {Chen}}, \bibinfo {author} {\bibfnamefont
  {H.}~\bibnamefont {Shi}}, \bibinfo {author} {\bibfnamefont {M.}~\bibnamefont
  {Motta}}, \bibinfo {author} {\bibfnamefont {C.}~\bibnamefont {Niu}}, \bibinfo
  {author} {\bibfnamefont {U.}~\bibnamefont {Ray}}, \bibinfo {author}
  {\bibfnamefont {S.}~\bibnamefont {Guo}}, \bibinfo {author} {\bibfnamefont
  {R.~J.}\ \bibnamefont {Anderson}}, \bibinfo {author} {\bibfnamefont
  {J.}~\bibnamefont {Li}}, \bibinfo {author} {\bibfnamefont {L.~N.}\
  \bibnamefont {Tran}}, \bibinfo {author} {\bibfnamefont {C.-N.}\ \bibnamefont
  {Yeh}}, \bibinfo {author} {\bibfnamefont {B.}~\bibnamefont {Mussard}},
  \bibinfo {author} {\bibfnamefont {S.}~\bibnamefont {Sharma}}, \bibinfo
  {author} {\bibfnamefont {F.}~\bibnamefont {Bruneval}}, \bibinfo {author}
  {\bibfnamefont {M.}~\bibnamefont {Van~Schilfgaarde}}, \bibinfo {author}
  {\bibfnamefont {G.~H.}\ \bibnamefont {Booth}}, \bibinfo {author}
  {\bibfnamefont {G.~K.-L.}\ \bibnamefont {Chan}}, \bibinfo {author}
  {\bibfnamefont {S.}~\bibnamefont {Zhang}}, \bibinfo {author} {\bibfnamefont
  {E.}~\bibnamefont {Gull}}, \bibinfo {author} {\bibfnamefont {D.}~\bibnamefont
  {Zgid}}, \bibinfo {author} {\bibfnamefont {A.}~\bibnamefont {Millis}},
  \bibinfo {author} {\bibfnamefont {C.~J.}\ \bibnamefont {Umrigar}}, \bibinfo
  {author} {\bibfnamefont {L.~K.}\ \bibnamefont {Wagner}}, \ and\ \bibinfo
  {author} {\bibnamefont {{Simons Collaboration on the Many-Electron
  Problem}}},\ }\href {\doibase 10.1103/PhysRevX.10.011041} {\bibfield
  {journal} {\bibinfo  {journal} {Physical Review X}\ }\textbf {\bibinfo
  {volume} {10}},\ \bibinfo {pages} {011041} (\bibinfo {year}
  {2020})}\BibitemShut {NoStop}%
\bibitem [{\citenamefont {Blunt}\ \emph {et~al.}(2015)\citenamefont {Blunt},
  \citenamefont {Alavi},\ and\ \citenamefont {Booth}}]{blunt_krylov_2015}%
  \BibitemOpen
  \bibfield  {author} {\bibinfo {author} {\bibfnamefont {N.~S.}\ \bibnamefont
  {Blunt}}, \bibinfo {author} {\bibfnamefont {A.}~\bibnamefont {Alavi}}, \ and\
  \bibinfo {author} {\bibfnamefont {G.~H.}\ \bibnamefont {Booth}},\ }\href
  {\doibase 10.1103/PhysRevLett.115.050603} {\bibfield  {journal} {\bibinfo
  {journal} {Phys. Rev. Lett.}\ }\textbf {\bibinfo {volume} {115}},\ \bibinfo
  {pages} {050603} (\bibinfo {year} {2015})}\BibitemShut {NoStop}%
\bibitem [{\citenamefont {Blunt}\ \emph {et~al.}(2017)\citenamefont {Blunt},
  \citenamefont {Booth},\ and\ \citenamefont {Alavi}}]{blunt_density_2017}%
  \BibitemOpen
  \bibfield  {author} {\bibinfo {author} {\bibfnamefont {N.~S.}\ \bibnamefont
  {Blunt}}, \bibinfo {author} {\bibfnamefont {G.~H.}\ \bibnamefont {Booth}}, \
  and\ \bibinfo {author} {\bibfnamefont {A.}~\bibnamefont {Alavi}},\ }\href
  {\doibase 10.1063/1.4986963} {\bibfield  {journal} {\bibinfo  {journal} {The
  Journal of Chemical Physics}\ }\textbf {\bibinfo {volume} {146}},\ \bibinfo
  {pages} {244105} (\bibinfo {year} {2017})},\ \Eprint
  {http://arxiv.org/abs/https://pubs.aip.org/aip/jcp/article-pdf/doi/10.1063/1.4986963/15527144/244105\_1\_online.pdf}
  {https://pubs.aip.org/aip/jcp/article-pdf/doi/10.1063/1.4986963/15527144/244105\_1\_online.pdf}
  \BibitemShut {NoStop}%
\bibitem [{\citenamefont {Choo}\ \emph {et~al.}(2018)\citenamefont {Choo},
  \citenamefont {Carleo}, \citenamefont {Regnault},\ and\ \citenamefont
  {Neupert}}]{choo_symmetries_2018}%
  \BibitemOpen
  \bibfield  {author} {\bibinfo {author} {\bibfnamefont {K.}~\bibnamefont
  {Choo}}, \bibinfo {author} {\bibfnamefont {G.}~\bibnamefont {Carleo}},
  \bibinfo {author} {\bibfnamefont {N.}~\bibnamefont {Regnault}}, \ and\
  \bibinfo {author} {\bibfnamefont {T.}~\bibnamefont {Neupert}},\ }\href
  {\doibase 10.1103/PhysRevLett.121.167204} {\bibfield  {journal} {\bibinfo
  {journal} {Phys. Rev. Lett.}\ }\textbf {\bibinfo {volume} {121}},\ \bibinfo
  {pages} {167204} (\bibinfo {year} {2018})}\BibitemShut {NoStop}%
\bibitem [{\citenamefont {Benavides-Riveros}\ \emph {et~al.}(2022)\citenamefont
  {Benavides-Riveros}, \citenamefont {Chen}, \citenamefont {Schilling},
  \citenamefont {Mantilla},\ and\ \citenamefont
  {Pittalis}}]{benavides_excitations_2022}%
  \BibitemOpen
  \bibfield  {author} {\bibinfo {author} {\bibfnamefont {C.~L.}\ \bibnamefont
  {Benavides-Riveros}}, \bibinfo {author} {\bibfnamefont {L.}~\bibnamefont
  {Chen}}, \bibinfo {author} {\bibfnamefont {C.}~\bibnamefont {Schilling}},
  \bibinfo {author} {\bibfnamefont {S.}~\bibnamefont {Mantilla}}, \ and\
  \bibinfo {author} {\bibfnamefont {S.}~\bibnamefont {Pittalis}},\ }\href
  {\doibase 10.1103/PhysRevLett.129.066401} {\bibfield  {journal} {\bibinfo
  {journal} {Phys. Rev. Lett.}\ }\textbf {\bibinfo {volume} {129}},\ \bibinfo
  {pages} {066401} (\bibinfo {year} {2022})}\BibitemShut {NoStop}%
\bibitem [{\citenamefont {Otis}\ and\ \citenamefont
  {Neuscamman}(2023{\natexlab{a}})}]{otis_promising_2023}%
  \BibitemOpen
  \bibfield  {author} {\bibinfo {author} {\bibfnamefont {L.}~\bibnamefont
  {Otis}}\ and\ \bibinfo {author} {\bibfnamefont {E.}~\bibnamefont
  {Neuscamman}},\ }\href {\doibase https://doi.org/10.1002/wcms.1659}
  {\bibfield  {journal} {\bibinfo  {journal} {WIREs Computational Molecular
  Science}\ }\textbf {\bibinfo {volume} {13}},\ \bibinfo {pages} {e1659}
  (\bibinfo {year} {2023}{\natexlab{a}})}\BibitemShut {NoStop}%
\bibitem [{\citenamefont {Filippi}\ \emph {et~al.}(2009)\citenamefont
  {Filippi}, \citenamefont {Zaccheddu},\ and\ \citenamefont
  {Buda}}]{filippi_absorption_2009}%
  \BibitemOpen
  \bibfield  {author} {\bibinfo {author} {\bibfnamefont {C.}~\bibnamefont
  {Filippi}}, \bibinfo {author} {\bibfnamefont {M.}~\bibnamefont {Zaccheddu}},
  \ and\ \bibinfo {author} {\bibfnamefont {F.}~\bibnamefont {Buda}},\ }\href
  {\doibase 10.1021/ct900227j} {\bibfield  {journal} {\bibinfo  {journal}
  {Journal of Chemical Theory and Computation}\ }\textbf {\bibinfo {volume}
  {5}},\ \bibinfo {pages} {2074} (\bibinfo {year} {2009})}\BibitemShut
  {NoStop}%
\bibitem [{\citenamefont {Zhao}\ and\ \citenamefont
  {Neuscamman}(2016)}]{zhao_efficient_2016}%
  \BibitemOpen
  \bibfield  {author} {\bibinfo {author} {\bibfnamefont {L.}~\bibnamefont
  {Zhao}}\ and\ \bibinfo {author} {\bibfnamefont {E.}~\bibnamefont
  {Neuscamman}},\ }\href {\doibase 10.1021/acs.jctc.6b00508} {\bibfield
  {journal} {\bibinfo  {journal} {Journal of Chemical Theory and Computation}\
  }\textbf {\bibinfo {volume} {12}},\ \bibinfo {pages} {3436} (\bibinfo {year}
  {2016})}\BibitemShut {NoStop}%
\bibitem [{\citenamefont {Shea}\ and\ \citenamefont
  {Neuscamman}(2017)}]{shea_size_2017}%
  \BibitemOpen
  \bibfield  {author} {\bibinfo {author} {\bibfnamefont {J.~A.~R.}\
  \bibnamefont {Shea}}\ and\ \bibinfo {author} {\bibfnamefont {E.}~\bibnamefont
  {Neuscamman}},\ }\href {\doibase 10.1021/acs.jctc.7b00923} {\bibfield
  {journal} {\bibinfo  {journal} {Journal of Chemical Theory and Computation}\
  }\textbf {\bibinfo {volume} {13}},\ \bibinfo {pages} {6078} (\bibinfo {year}
  {2017})}\BibitemShut {NoStop}%
\bibitem [{\citenamefont {Pineda~Flores}\ and\ \citenamefont
  {Neuscamman}(2019)}]{pineda_flores_excited_2019}%
  \BibitemOpen
  \bibfield  {author} {\bibinfo {author} {\bibfnamefont {S.~D.}\ \bibnamefont
  {Pineda~Flores}}\ and\ \bibinfo {author} {\bibfnamefont {E.}~\bibnamefont
  {Neuscamman}},\ }\href {\doibase 10.1021/acs.jpca.8b10671} {\bibfield
  {journal} {\bibinfo  {journal} {The Journal of Physical Chemistry A}\
  }\textbf {\bibinfo {volume} {123}},\ \bibinfo {pages} {1487} (\bibinfo {year}
  {2019})}\BibitemShut {NoStop}%
\bibitem [{\citenamefont {Otis}\ and\ \citenamefont
  {Neuscamman}(2023{\natexlab{b}})}]{otis_optimization_2023}%
  \BibitemOpen
  \bibfield  {author} {\bibinfo {author} {\bibfnamefont {L.}~\bibnamefont
  {Otis}}\ and\ \bibinfo {author} {\bibfnamefont {E.}~\bibnamefont
  {Neuscamman}},\ }\href {\doibase 10.1021/acs.jctc.2c00642} {\bibfield
  {journal} {\bibinfo  {journal} {Journal of Chemical Theory and Computation}\
  }\textbf {\bibinfo {volume} {19}},\ \bibinfo {pages} {767} (\bibinfo {year}
  {2023}{\natexlab{b}})}\BibitemShut {NoStop}%
\bibitem [{\citenamefont {Cuzzocrea}\ \emph {et~al.}(2020)\citenamefont
  {Cuzzocrea}, \citenamefont {Scemama}, \citenamefont {Briels}, \citenamefont
  {Moroni},\ and\ \citenamefont {Filippi}}]{cuzzocrea_variational_2020}%
  \BibitemOpen
  \bibfield  {author} {\bibinfo {author} {\bibfnamefont {A.}~\bibnamefont
  {Cuzzocrea}}, \bibinfo {author} {\bibfnamefont {A.}~\bibnamefont {Scemama}},
  \bibinfo {author} {\bibfnamefont {W.~J.}\ \bibnamefont {Briels}}, \bibinfo
  {author} {\bibfnamefont {S.}~\bibnamefont {Moroni}}, \ and\ \bibinfo {author}
  {\bibfnamefont {C.}~\bibnamefont {Filippi}},\ }\href {\doibase
  10.1021/acs.jctc.0c00147} {\bibfield  {journal} {\bibinfo  {journal} {Journal
  of Chemical Theory and Computation}\ }\textbf {\bibinfo {volume} {16}},\
  \bibinfo {pages} {4203} (\bibinfo {year} {2020})}\BibitemShut {NoStop}%
\bibitem [{\citenamefont {Higgott}\ \emph {et~al.}(2019)\citenamefont
  {Higgott}, \citenamefont {Wang},\ and\ \citenamefont
  {Brierley}}]{Higgott2019variationalquantum}%
  \BibitemOpen
  \bibfield  {author} {\bibinfo {author} {\bibfnamefont {O.}~\bibnamefont
  {Higgott}}, \bibinfo {author} {\bibfnamefont {D.}~\bibnamefont {Wang}}, \
  and\ \bibinfo {author} {\bibfnamefont {S.}~\bibnamefont {Brierley}},\ }\href
  {\doibase 10.22331/q-2019-07-01-156} {\bibfield  {journal} {\bibinfo
  {journal} {{Quantum}}\ }\textbf {\bibinfo {volume} {3}},\ \bibinfo {pages}
  {156} (\bibinfo {year} {2019})}\BibitemShut {NoStop}%
\bibitem [{\citenamefont {Pathak}\ \emph {et~al.}(2021)\citenamefont {Pathak},
  \citenamefont {Busemeyer}, \citenamefont {Rodrigues},\ and\ \citenamefont
  {Wagner}}]{pathak_excited_2021}%
  \BibitemOpen
  \bibfield  {author} {\bibinfo {author} {\bibfnamefont {S.}~\bibnamefont
  {Pathak}}, \bibinfo {author} {\bibfnamefont {B.}~\bibnamefont {Busemeyer}},
  \bibinfo {author} {\bibfnamefont {J.~N.~B.}\ \bibnamefont {Rodrigues}}, \
  and\ \bibinfo {author} {\bibfnamefont {L.~K.}\ \bibnamefont {Wagner}},\
  }\href {\doibase 10.1063/5.0030949} {\bibfield  {journal} {\bibinfo
  {journal} {The Journal of Chemical Physics}\ }\textbf {\bibinfo {volume}
  {154}},\ \bibinfo {pages} {034101} (\bibinfo {year} {2021})}\BibitemShut
  {NoStop}%
\bibitem [{\citenamefont {Shepard}\ \emph {et~al.}(2022)\citenamefont
  {Shepard}, \citenamefont {Panadés-Barrueta}, \citenamefont {Moroni},
  \citenamefont {Scemama},\ and\ \citenamefont
  {Filippi}}]{shepard_double_2022}%
  \BibitemOpen
  \bibfield  {author} {\bibinfo {author} {\bibfnamefont {S.}~\bibnamefont
  {Shepard}}, \bibinfo {author} {\bibfnamefont {R.~L.}\ \bibnamefont
  {Panadés-Barrueta}}, \bibinfo {author} {\bibfnamefont {S.}~\bibnamefont
  {Moroni}}, \bibinfo {author} {\bibfnamefont {A.}~\bibnamefont {Scemama}}, \
  and\ \bibinfo {author} {\bibfnamefont {C.}~\bibnamefont {Filippi}},\ }\href
  {\doibase 10.1021/acs.jctc.2c00769} {\bibfield  {journal} {\bibinfo
  {journal} {Journal of Chemical Theory and Computation}\ }\textbf {\bibinfo
  {volume} {18}},\ \bibinfo {pages} {6722} (\bibinfo {year}
  {2022})}\BibitemShut {NoStop}%
\bibitem [{\citenamefont {Entwistle}\ \emph {et~al.}(2023)\citenamefont
  {Entwistle}, \citenamefont {Sch{\"{a}}tzle}, \citenamefont {Erdman},
  \citenamefont {Hermann},\ and\ \citenamefont
  {No{\'e}}}]{entwistle_electronic_2023}%
  \BibitemOpen
  \bibfield  {author} {\bibinfo {author} {\bibfnamefont {M.~T.}\ \bibnamefont
  {Entwistle}}, \bibinfo {author} {\bibfnamefont {Z.}~\bibnamefont
  {Sch{\"{a}}tzle}}, \bibinfo {author} {\bibfnamefont {P.~A.}\ \bibnamefont
  {Erdman}}, \bibinfo {author} {\bibfnamefont {J.}~\bibnamefont {Hermann}}, \
  and\ \bibinfo {author} {\bibfnamefont {F.}~\bibnamefont {No{\'e}}},\ }\href
  {\doibase 10.1038/s41467-022-35534-5} {\bibfield  {journal} {\bibinfo
  {journal} {Nature Communications}\ }\textbf {\bibinfo {volume} {14}},\
  \bibinfo {pages} {274} (\bibinfo {year} {2023})}\BibitemShut {NoStop}%
\bibitem [{\citenamefont {Pfau}\ \emph {et~al.}(2023)\citenamefont {Pfau},
  \citenamefont {Axelrod}, \citenamefont {Sutterud}, \citenamefont {von
  Glehn},\ and\ \citenamefont {Spencer}}]{pfau_natural_2023}%
  \BibitemOpen
  \bibfield  {author} {\bibinfo {author} {\bibfnamefont {D.}~\bibnamefont
  {Pfau}}, \bibinfo {author} {\bibfnamefont {S.}~\bibnamefont {Axelrod}},
  \bibinfo {author} {\bibfnamefont {H.}~\bibnamefont {Sutterud}}, \bibinfo
  {author} {\bibfnamefont {I.}~\bibnamefont {von Glehn}}, \ and\ \bibinfo
  {author} {\bibfnamefont {J.~S.}\ \bibnamefont {Spencer}},\ }\href
  {https://arxiv.org/abs/2308.16848v1} {\bibfield  {journal} {\bibinfo
  {journal} {arXiv.org}\ } (\bibinfo {year} {2023})}\BibitemShut {NoStop}%
\bibitem [{\citenamefont {Theophilou}(1979)}]{theophilou_energy_1979}%
  \BibitemOpen
  \bibfield  {author} {\bibinfo {author} {\bibfnamefont {A.~K.}\ \bibnamefont
  {Theophilou}},\ }\href {\doibase 10.1088/0022-3719/12/24/013} {\bibfield
  {journal} {\bibinfo  {journal} {Journal of Physics C: Solid State Physics}\
  }\textbf {\bibinfo {volume} {12}},\ \bibinfo {pages} {5419} (\bibinfo {year}
  {1979})}\BibitemShut {NoStop}%
\bibitem [{\citenamefont {Gross}\ \emph {et~al.}(1988)\citenamefont {Gross},
  \citenamefont {Oliveira},\ and\ \citenamefont {Kohn}}]{gross_rayleigh_1988}%
  \BibitemOpen
  \bibfield  {author} {\bibinfo {author} {\bibfnamefont {E.~K.~U.}\
  \bibnamefont {Gross}}, \bibinfo {author} {\bibfnamefont {L.~N.}\ \bibnamefont
  {Oliveira}}, \ and\ \bibinfo {author} {\bibfnamefont {W.}~\bibnamefont
  {Kohn}},\ }\href {\doibase 10.1103/PhysRevA.37.2805} {\bibfield  {journal}
  {\bibinfo  {journal} {Phys. Rev. A}\ }\textbf {\bibinfo {volume} {37}},\
  \bibinfo {pages} {2805} (\bibinfo {year} {1988})}\BibitemShut {NoStop}%
\bibitem [{\citenamefont {Amari}(1998)}]{natural_gradient}%
  \BibitemOpen
  \bibfield  {author} {\bibinfo {author} {\bibfnamefont {S.-i.}\ \bibnamefont
  {Amari}},\ }\href {\doibase 10.1162/089976698300017746} {\bibfield  {journal}
  {\bibinfo  {journal} {Neural Computation}\ }\textbf {\bibinfo {volume}
  {10}},\ \bibinfo {pages} {251} (\bibinfo {year} {1998})}\BibitemShut
  {NoStop}%
\bibitem [{\citenamefont {Sorella}(1998)}]{sorella_green_1998}%
  \BibitemOpen
  \bibfield  {author} {\bibinfo {author} {\bibfnamefont {S.}~\bibnamefont
  {Sorella}},\ }\href {\doibase 10.1103/PhysRevLett.80.4558} {\bibfield
  {journal} {\bibinfo  {journal} {Phys. Rev. Lett.}\ }\textbf {\bibinfo
  {volume} {80}},\ \bibinfo {pages} {4558} (\bibinfo {year}
  {1998})}\BibitemShut {NoStop}%
\bibitem [{\citenamefont {Wheeler}\ \emph {et~al.}(2023)\citenamefont
  {Wheeler}, \citenamefont {Pathak}, \citenamefont {Kleiner}, \citenamefont
  {Yuan}, \citenamefont {Rodrigues}, \citenamefont {Lorsung}, \citenamefont
  {Krongchon}, \citenamefont {Chang}, \citenamefont {Zhou}, \citenamefont
  {Busemeyer}, \citenamefont {Williams}, \citenamefont {Muñoz}, \citenamefont
  {Chow},\ and\ \citenamefont {Wagner}}]{pyqmc}%
  \BibitemOpen
  \bibfield  {author} {\bibinfo {author} {\bibfnamefont {W.~A.}\ \bibnamefont
  {Wheeler}}, \bibinfo {author} {\bibfnamefont {S.}~\bibnamefont {Pathak}},
  \bibinfo {author} {\bibfnamefont {K.~G.}\ \bibnamefont {Kleiner}}, \bibinfo
  {author} {\bibfnamefont {S.}~\bibnamefont {Yuan}}, \bibinfo {author}
  {\bibfnamefont {J.~N.~B.}\ \bibnamefont {Rodrigues}}, \bibinfo {author}
  {\bibfnamefont {C.}~\bibnamefont {Lorsung}}, \bibinfo {author} {\bibfnamefont
  {K.}~\bibnamefont {Krongchon}}, \bibinfo {author} {\bibfnamefont
  {Y.}~\bibnamefont {Chang}}, \bibinfo {author} {\bibfnamefont
  {Y.}~\bibnamefont {Zhou}}, \bibinfo {author} {\bibfnamefont {B.}~\bibnamefont
  {Busemeyer}}, \bibinfo {author} {\bibfnamefont {K.~T.}\ \bibnamefont
  {Williams}}, \bibinfo {author} {\bibfnamefont {A.}~\bibnamefont {Muñoz}},
  \bibinfo {author} {\bibfnamefont {C.~Y.}\ \bibnamefont {Chow}}, \ and\
  \bibinfo {author} {\bibfnamefont {L.~K.}\ \bibnamefont {Wagner}},\ }\href
  {\doibase 10.1063/5.0139024} {\bibfield  {journal} {\bibinfo  {journal} {The
  Journal of Chemical Physics}\ }\textbf {\bibinfo {volume} {158}},\ \bibinfo
  {pages} {114801} (\bibinfo {year} {2023})},\ \Eprint
  {http://arxiv.org/abs/https://pubs.aip.org/aip/jcp/article-pdf/doi/10.1063/5.0139024/16794896/114801\_1\_online.pdf}
  {https://pubs.aip.org/aip/jcp/article-pdf/doi/10.1063/5.0139024/16794896/114801\_1\_online.pdf}
  \BibitemShut {NoStop}%
\bibitem [{\citenamefont {Tobias}\ \emph {et~al.}(1960)\citenamefont {Tobias},
  \citenamefont {Fallon},\ and\ \citenamefont
  {Vanderslice}}]{tobiasPotentialEnergyCurves1960}%
  \BibitemOpen
  \bibfield  {author} {\bibinfo {author} {\bibfnamefont {I.}~\bibnamefont
  {Tobias}}, \bibinfo {author} {\bibfnamefont {R.~J.}\ \bibnamefont {Fallon}},
  \ and\ \bibinfo {author} {\bibfnamefont {J.~T.}\ \bibnamefont
  {Vanderslice}},\ }\href {\doibase 10.1063/1.1731475} {\bibfield  {journal}
  {\bibinfo  {journal} {The Journal of Chemical Physics}\ }\textbf {\bibinfo
  {volume} {33}},\ \bibinfo {pages} {1638} (\bibinfo {year}
  {1960})}\BibitemShut {NoStop}%
\bibitem [{\citenamefont {Bennett}\ \emph {et~al.}(2017)\citenamefont
  {Bennett}, \citenamefont {Melton}, \citenamefont {Annaberdiyev},
  \citenamefont {Wang}, \citenamefont {Shulenburger},\ and\ \citenamefont
  {Mitas}}]{bennet_new_2017}%
  \BibitemOpen
  \bibfield  {author} {\bibinfo {author} {\bibfnamefont {M.~C.}\ \bibnamefont
  {Bennett}}, \bibinfo {author} {\bibfnamefont {C.~A.}\ \bibnamefont {Melton}},
  \bibinfo {author} {\bibfnamefont {A.}~\bibnamefont {Annaberdiyev}}, \bibinfo
  {author} {\bibfnamefont {G.}~\bibnamefont {Wang}}, \bibinfo {author}
  {\bibfnamefont {L.}~\bibnamefont {Shulenburger}}, \ and\ \bibinfo {author}
  {\bibfnamefont {L.}~\bibnamefont {Mitas}},\ }\href {\doibase
  10.1063/1.4995643} {\bibfield  {journal} {\bibinfo  {journal} {The Journal of
  Chemical Physics}\ }\textbf {\bibinfo {volume} {147}},\ \bibinfo {pages}
  {224106} (\bibinfo {year} {2017})}\BibitemShut {NoStop}%
\end{thebibliography}%

\end{document}